\documentclass[showpacs,preprint,aps]{revtex4}
\usepackage{graphicx}% Include figure files
\usepackage{dcolumn}% Align table columns on decimal point
\usepackage{bm}% bold math
\begin{document}
\setcounter{page}{1}
\title
{Critical view of WKB decay widths}
\author
{N. G. Kelkar and H. M. Casta\~neda}
\affiliation{
Departamento de Fisica, Universidad de los Andes,
Cra.1E No.18A-10, Bogota, Colombia}
\begin{abstract}
A detailed comparison of the expressions for the decay widths obtained 
within the semiclassical WKB approximation using different approaches
to the tunneling problem is performed. The differences between the 
available 
improved formulae for tunneling near the top and the bottom of the barrier
are investigated. Though the simple
WKB method gives the right order of magnitude of the decay widths, 
a small number of parameters are often fitted. The need to perform the 
fitting procedure remaining consistently within the WKB framework is 
emphasized in the context of the fission model based calculations. 
Calculations for the decay widths of some recently found super heavy 
nuclei using microscopic alpha-nucleus potentials are presented to 
demonstrate the importance of a consistent WKB calculation. 
The half-lives 
are found to be sensitive to the density dependence of the nucleon-nucleon 
interaction and the implementation of the Bohr-Sommerfeld quantization 
condition inherent in the WKB approach.
\end{abstract}
\maketitle

\section{Introduction}
The Wentzel-Kramers-Brillouin or the WKB approximation \cite{mrwkbs, berry}, 
sometimes
also known as the BWK \cite{kemble}, the semiclassical approximation or the
phase integral method \cite{froeman, froeman2} has been widely used in the 
evaluation of the half-lives of radioactive nuclei. It was evident from the 
historical papers of Gamow \cite{gamow} and Condon and Gurney \cite{condon} 
that one could treat the alpha decay of nuclei in terms of the tunneling of a 
preformed $\alpha$-particle confined to the interior of the nucleus, 
through the Coulomb potential barrier of the alpha-nucleus system. The WKB 
approximation which is really applicable when a problem can be reduced to a 
one-dimensional one was found suitable to evaluate the barrier penetration 
probabilities and the decay width in general was defined as a product of 
the frequency of collisions of the $\alpha$ with the barrier (the so-called
assault frequency) and the penetration probability. The objective of the
present work is to critically examine the decay widths obtained within 
some entirely different approaches to the tunneling problem, however, all
working within a WKB framework. To be specific, we examine four 
approaches; the first a two potential approach (TPA) \cite{gurvitz}, another a 
path integral method with Jost functions \cite{froeman}, a third 
one using comparison equations to obtain improved WKB formulae \cite{shepard} 
and a fourth one involving a super asymmetric fission model (SAFM) 
\cite{poenaru}. 
It is gratifying to know that all the
four approaches indeed lead to the same formulae for the WKB decay widths. 
A WKB expression for the vibrational energy, $E_{\nu}$, emerges from the 
above comparisons. For the first three approaches, the vibrational energy 
(and hence the assault frequency) can be simply evaluated from 
the potential and the tunneling particle energy. 
In the SAFM however, the equation for $E_{\nu}$ turns out to 
be a transcendental equation. 
However, this equation is often neglected by the SAFM calculations 
in literature and a fit to the vibrational energies is performed, 
leaving the calculation somewhat incompatible with the WKB framework. 

Besides this, the widths 
calculated within the SAFM often neglect the Bohr-Sommerfeld condition and 
the Langer modification which are essential ingredients of the WKB framework. 
The Langer modification is a necessary transformation while going from 
the one dimensional problem with $x$ ranging from $-\infty \,\rightarrow \, 
\infty$ to the radial one-dimensional tunneling problem with $r$ 
ranging from $0 \rightarrow \infty$.
To demonstrate the 
importance of performing a completely consistent WKB calculation of widths, we
perform a realistic calculation of the alpha decay widths of super heavy 
nuclei which are a topic of current interest.

With most radioactive decays occurring away from the extremes of the potential
barrier, the standard WKB is found to be a reasonable approximation for such 
calculations. However, for specific cases, where the tunneling can occur 
near the top or the bottom of the barrier, the validity of this approach 
becomes questionable. Some attempts at obtaining improved formulae which can 
be used near the extremes of the barrier do exist \cite{shepard,froeman} and 
will be investigated by applying to realistic examples in the present work.

The alpha - daughter nucleus interaction in the present work is described 
using folding model potentials with a realistic nucleon-nucleon 
(NN) interaction 
\cite{satchler}. Such a model has been shown to be quite successful in 
predicting half-lives of unstable nuclei \cite{chin}. 
We perform calculations with the added ingredient of a density dependent 
NN interaction and find the results and fitted parameters sensitive to
this input. In the next section, we perform a comparison of the different 
WKB approaches as mentioned above. In Section III, we briefly present the 
relevant formulae of the potentials used. The results are discussed in 
Section IV. 
 
\section{The quasiclassical alpha tunneling problem}
Starting with the description of a radioactive nucleus as a cluster of 
its daughter nucleus and an alpha particle, it is by now standard practice 
to study the alpha decay of nuclei as a tunneling of the $\alpha$ through 
the potential barrier of the alpha-daughter nucleus system. Though most 
quasiclassical approaches agree on the proportionality of the decay width 
to an exponential factor, namely, $\Gamma \propto e^{-2G}$, where $G$ is 
the famous Gamow factor, the details of the calculations vary depending on
the approach used. 
\begin{figure}[ht]
\includegraphics[width=8cm,height=5cm]{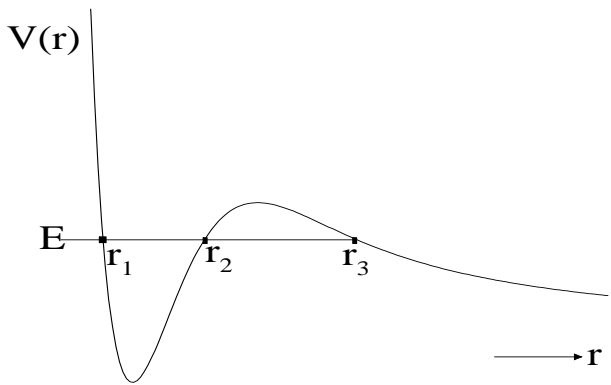}
\caption{\label{fig:eps1}Typical potential for an alpha-nucleus tunneling
problem.}
\end{figure}
Typically, one considers the tunneling of the $\alpha$ 
through an $r$-space potential of the form, 
\begin{equation}\label{pot1}
V(r)\, =\, V_n(r)\,+\,V_c(r)\,+\,{\hbar^2\,(l\,+\,1/2)^2 \over \mu\,r^2}\, ,
\end{equation} 
where $V_n(r)$ and $V_c(r)$ are the nuclear and Coulomb parts of the 
$\alpha$-nucleus (daughter) potential, $r$ the distance between
the centres of mass of the daughter nucleus and alpha and $\mu$ 
their reduced mass. The last term 
represents the Langer modified centrifugal barrier \cite{langer}. With the 
WKB being valid for one-dimensional problems, the above modification from 
$l(l+1) \,\rightarrow \,(l+1/2)^2$ is essential to ensure the correct 
behaviour of the WKB scattered radial wave function near the origin as well as
the validity of the connection formulas used \cite{morehead}. 
Another 
requisite for the correct use of the WKB method is the Bohr-Sommerfeld 
quantization condition, which for an alpha with energy $E$ is given as,  
\begin{equation}\label{bohrsomer}
\int_{r_1}^{r_2} \,\, K(r)\,dr\,=\,(n\,+\,1/2)\,\pi
\end{equation}
where $K(r) \,=\, \sqrt{{2\mu \over \hbar^2}\,(|V(r)\,-\,E)|}$, $n$ is the 
number of nodes of the quasibound wave function of $\alpha$-nucleus 
relative motion and $r_1$ and $r_2$ which are solutions of $V(r) \,=\,E$, are
the classical turning points (as shown in Fig. 1). We shall now examine the
decay widths obtained within different WKB based approaches to the above 
problem. 
\subsection{The two potential approach}
Starting with a typical potential as in Fig. 1, the authors in 
\cite{gurvitz} consider the tunneling problem of a metastable 
(quasistationary) state and obtain a perturbative expansion for the decay 
width of the metastable state. The potential is split into two parts, 
$V(r)\,=\,U(r)\,+\,W(r)$, where the authors first consider an unperturbed 
bound wave function $\Phi_0(r)$, which is an eigenstate of the Hamiltonian, 
$H_0 \,=\, -(\hbar^2/2\mu)\nabla^2\,+\,U(r)$. $W(r)$ is a perturbation and 
when it is switched on, $\Phi_0(r)$ is not an eigenfunction of 
$H\,=\,H_0\,+\,W(r)$, but one rather has a wave packet which is expressed 
as an expansion in terms of $\Phi_0(r)$ and the continuum wave functions,  
$\Phi_k(r)$. Once the perturbative expansion for the width is obtained in 
terms of the wave functions, retaining the first term and expressing the 
wave function using the semiclassical WKB approximation, the width is given 
as,
\begin{equation}\label{gurwidth}
\Gamma_{TPA}(E)\,=\,{\hbar^2 \over 2 \,\mu}\,\,\biggl[\,\int_{r_1}^{r_2}\,
{dr \over k(r)}\,\biggr]^{-1}\,e^{-2\int_{r_2}^{r_3}\,k(r)\,dr}
\end{equation} 
where, $k(r) \,=\, \sqrt{{2\mu \over \hbar^2}\,(|V(r)\,-\,E|)}$. The factor in 
front of the exponential arises from the normalization of the bound state 
wave function in the region between $r_1$ and $r_2$. Indeed, this factor is 
related to the so-called `assault frequency' of the tunneling particle at 
the barrier. Expressing the time interval, $\Delta t$, for the particle  
traversing a distance, $\Delta r$ as, 
\begin{equation}\label{2}
\Delta t \,=\, {\Delta r \over v(r)}\,=\,{\mu \, \Delta r \over \hbar\,k(r)}\,,
\end{equation}
the assault frequency $\nu$ can be written as the inverse of the time required 
to traverse the distance back and forth between the turning points $r_1$ and 
$r_2$ as \cite{froeman},
\begin{equation}\label{period}
\nu\,=\,T^{-1}\,=\,{\hbar \over 2\,\mu}\,\biggl[\,\int_{r_1}^{r_2}\, 
{dr \over 
\sqrt{{2\mu \over \hbar^2}\,(|V(r)\,-\,E|)}
}\,\biggr]^{-1}\,\,.
\end{equation}
Thus, $\Gamma_{TPA}(E)\,=\,\hbar\,\nu\,e^{-2W}$, where, $W\,=\,
\int_{r_2}^{r_3}\,k(r)\,dr$. It is interesting to note that the above 
definition of the period $T$ (which is twice the time spent between the 
turning points $r_1$ and $r_2$), 
is formally similar to that of ``traversal time in tunneling" 
as defined by B\"uttiker and Landauer \cite{buttik}. They defined the 
time for a particle traversing the barrier and hence obtained a similar 
definition as that for $T/2$ above, but within the limits $r_2$ to $r_3$. 
\subsection{Improved WKB widths}
The width obtained in the two potential approach is in accord with the 
widths obtained in \cite{froeman} and \cite{shepard} for energies well away
from the top or the bottom of the barrier. For example, in \cite{shepard}, 
considering a double humped (DH) barrier in one dimension (with cartesian 
coordinates) and 
using the method of comparison equations, the decay width for all energies 
except near the very top or the very bottom of the barrier was found to be, 
\begin{equation}\label{shepwidth}
\Gamma_{DH}(E)\,=\,2\,\hbar\,\nu_{DH}\,\rm{ln}[\,1\,+\,e^{-2W}\,]\, 
\simeq\,2\,\hbar\,\nu_{DH}\,e^{-2W}
\end{equation}
where, $\nu_{DH}$ is the assault frequency in the case of a double humped 
barrier. Replacing $\nu_{DH}\,=\,\nu/2$, for small values of the exponent, 
$\Gamma_{DH}(E)$ indeed reduces exactly to $\Gamma_{TPA}(E)$. 
\subsubsection{At the base of the barrier}
The width for energies of the tunneling particle near the bottom of the well 
was found in \cite{shepard} to be,
\begin{equation}\label{shepwidthlow}
\Gamma_{DH}^{low}(E_n)\,=\,2\,\hbar\,\nu_{DH}\,\rm{ln}[\,1\,+\,\alpha^{-1}\,
e^{-2W}]\,\simeq\,\alpha^{-1}\,\Gamma_{DH}(E_n)\,,
\end{equation}
where, $\alpha^{-1} \,=\,(1/n!)\,\sqrt{2\pi}\,[(n+1/2)/e]^{n+1/2}$. The 
relation obtained in \cite{gurvitz} using the TPA is somewhat different,
namely, 
\begin{equation}\label{gurwidthlow}
\Gamma_{TPA}^{low}(E_n)\,=\,{\sqrt{2} \over \pi}\,(\Gamma(1/4))^2\,(n+1/2)^{1/2}\,
\Gamma_{TPA}(E_n)\, .
\end{equation}
It is mentioned in \cite{gurvitz} that the above result coincides with that in 
\cite{hanggi} obtained using complex-time path integral methods \cite{weiss}. 
The prefactors in front of the exponential factors in the expression for
the widths in \cite{hanggi,weiss} depend on characteristic parameters of 
instanton trajectories, however, the above modification in (\ref{gurwidthlow}) 
does agree in form with that obtained in \cite{hanggi}. 
Though the TPA approach and that of comparison equations in \cite{shepard} 
do give the same expressions for the widths away from the extremes of the 
barrier ($\Gamma_{TPA}(E) = \Gamma_{DH}(E)$ was seen above), they
do not seem to agree on the situation at low energies. 
Eqs (\ref{shepwidthlow}) and (\ref{gurwidthlow}) seem to give 
very different results. For example, for $n=0,1$ respectively, one finds 
that $\Gamma_{DH}^{low}(E) = 1.075 \Gamma_{DH}(E)$ and $\Gamma_{DH}^{low}(E) 
= 1.027 \Gamma_{DH}(E)$, 
whereas, $\Gamma_{TPA}^{low}(E) = 4.184 \Gamma_{TPA}(E)$ and 
$\Gamma_{TPA}^{low}(E) = 7.247 \Gamma_{TPA}(E)$ respectively. 
In \cite{friedrich}, one can find yet another expression for tunneling 
near the bottom of the barrier such that the tunneling probability at the base
of the barrier vanishes exactly. 
\subsubsection{At the top of the barrier}
In \cite{shepard}, the decay width within the WKB approximation at the top of
the barrier was also evaluated and found to be,
\begin{equation}\label{sheptop}
\Gamma_{DH}^{top}(E)\,=\, {2 \,\hbar \, {\rm ln}\,[\,1\,+\,e^{-2W}\,] 
\over \biggl [\, T_{DH}\, -\, 2\,\hbar\,{d\phi \over dE}\,\biggr]}\, ,
\end{equation}
where, $T_{DH}$ is the period for moving back and forth between the two humps
and $\phi \,=\,arg\Gamma({1\over 2}\,-\,iW/\pi)\,+\,{W\over\pi}\,[\,
\rm{ln} (W/\pi)\,-\,1]$. A very similar formula was also found in 
\cite{froeman}, however, with a difference of a sign in the denominator. 
Having derived the expressions for the Jost function of a radial 
barrier transmission problem (with one hump) by the path integral method, 
the authors obtain the following expression for evaluating the width at the 
top of the barrier:
\begin{equation}\label{froformula}
\Gamma^{top}_{Fr\ddot{o}man}(E)\,=\,{2 \hbar\,[\,(1\,+\,e^{-2W})^{1/4}\,-\,
(1\,+\,e^{-2W})^{-1/4}\,] \over 
\biggl [\, T\, -\, 2\,\hbar\,{d\sigma \over dE}\,\biggr]}\, .
\end{equation}
The numerators in (\ref{froformula}) and (\ref{sheptop}) for energies 
near the top of the barrier are almost equal, i.e.,  
$2 \hbar\,[\,(1\,+\,e^{-2W})^{1/4}\,-\,
(1\,+\,e^{-2W})^{-1/4}\,]\,\simeq\,\hbar \rm{ln}\,[\,1\,+\,e^{-2W}\,]$. 
With the period, $T_{DH}$ being twice as much as $T$ and the factor 
$\phi\, = \,- 2\,\sigma$, Eqs (\ref{sheptop}) and (\ref{froformula}) indeed
agree but up to a sign in the denominator. 
In Eq. (5.7) of \cite{froeman}, however, one can see that there exists a 
choice for the sign appearing in front of $d\sigma / dE$ and the authors 
choose the negative sign without any particular justification. 
We shall later notice with a realistic example near the top of the barrier, 
that it is indeed the choice of a positive sign in the denominator 
which improves the WKB width estimate. The choice of a positive sign also 
brings (\ref{froformula}) in agreement with (\ref{sheptop}). 

We now move on to discuss one more approach which is popularly used in 
literature and has off late been often used for the evaluation of the 
half-lives of superheavy nuclei.
\subsection{Fission model approach}
An approach where the alpha decay is considered as a very asymmetric fission 
process was introduced about two decades ago by Poenaru and co-workers 
\cite{poenaru}. This model, also known as the super asymmetric fission 
model (SAFM) has recently been implemented extensively for the evaluation 
of the WKB widths of superheavy nuclei \cite{fairy,others}. The decay width
of a metastable state in the SAFM within the WKB framework is given as,
\begin{equation}\label{safm}
\Gamma_{SAFM}(E)\,=\,\nu\,P\,=\, {\,E_{\nu} \over \pi}\,\biggl 
(\,1\,+\,e^{2K}\,\biggr)^{-1}\, ,
\end{equation} 
where, $P$ is the probability of penetration through the potential barrier, 
$E_{\nu}$ is the ``vibrational energy", $ K \,=\,\int_{r_2}^{r_3}\,
\kappa(r)\,dr$ with $\kappa (r)\,=\,\sqrt{(2\mu/\hbar^2)\,(V(r) \,-\,
Q^{\prime})}$ and $Q^{\prime}\,=\,E\,+\,E_{\nu}$. The Gamow factor with $K$ 
over here differs from the $W$ occurring in the equations so far as obtained
in \cite{froeman,gurvitz,shepard} due the energy $E$ of the tunneling 
particle being replaced by $E \,+\,E_{\nu}$. Replacing the assault frequency 
$\nu$ from Eq. (\ref{period}) into the above Eq. (\ref{safm}), we indeed 
recover an expression similar to that of $\Gamma_{TPA}(E)$, with the $k(r)$ 
replaced here by $\kappa (r)$. The SAFM width is, 
\begin{equation}\label{safm2}
\Gamma_{SAFM}(E) \,=\, \,{\hbar^2 \over 2 \,\mu}\,\,\biggl[\,\int_{r_1}^{r_2}\,
{dr \over \kappa(r)}\,\biggr]^{-1}\,e^{-2\int_{r_2}^{r_3}\,\kappa(r)\,dr}
\end{equation}
where we have approximated $(\,1\,+\,e^{2K}\,)^{-1}\,\simeq 
\,e^{-2K}$ for sufficiently large $K$. In fact, if we start with the 
definition of $E_{\nu} \,=\,(1/2)\,\hbar\,\omega \,=\, (1/2)\,\hbar\,
(2\,\pi\,\nu)$ (as defined in the SAFM based works 
\cite{poenaru,fairy,others}) and use Eq. (\ref{period}) for $\nu\,=\,T^{-1}$, 
we obtain a theoretical relation for $E_{\nu}$, namely, 
\begin{equation}\label{vibenergy}
E_{\nu}\,=\,{\hbar^2 \,\pi \over 2\,\mu}\,\,\biggl [\,\int_{r_1}^{r_2}\, 
{dr \over \sqrt{(2\mu/\hbar^2)\,(V(r) \,-\,E\,-\,E_{\nu})}
} \,\biggr ]^{-1}\,\,.
\end{equation}
One could have of course inferred the above equation (\ref{vibenergy}) 
simply by the comparison of Eqs. (\ref{safm}) and (\ref{safm2}). 
Provided the potential is known, for a given energy $E$ of the tunneling 
particle, Eq. (\ref{vibenergy}) is a transcendental equation for the 
vibrational energy $E_{\nu}$.

Comparing the expressions for the widths, 
$\Gamma_{TPA}$, $\Gamma_{DH}$ and $\Gamma_{Fr\ddot{o}man}$, away from the extremes 
of the barrier (taken in the limit of large $W$ 
and negligible $d\sigma/dE$), one can see that indeed,  
\begin{equation}\label{compare3}
\Gamma_{TPA}(E)\,=\,\Gamma_{DH}(E)\,=\,\Gamma_{Fr\ddot{o}man}(E)\,=\, 
{\hbar^2 \over 2 \,\mu}\,\,\biggl[\,\int_{r_1}^{r_2}\,
{dr \over k(r)}\,\biggr]^{-1}\,e^{-2\int_{r_2}^{r_3}\,k(r)\,dr}\,.
\end{equation}
$\Gamma_{SAFM}$ agrees exactly in 
form with the above formulae for widths. 
The only difference lies in the replacement of $E$ by $E\,+\,E_{\nu}$ 
as mentioned before. 
It is interesting to note that even though the 
prefactor in front of the exponential 
in $\Gamma_{TPA}$ arises due to the 
normalization of the WKB wave function, it agrees exactly with the prefactors
in $\Gamma_{DH}$ and $\Gamma_{Fr\ddot{o}man}$ where it arises due to the 
replacement of the assault frequency as in (\ref{period}). 

Coming back to Eq. (\ref{vibenergy}), one notices that for a given potential, 
$V(r)$, in a particular decay problem with a given $Q$-value, $E_{\nu}$ can 
be determined by resolving Eq. (\ref{vibenergy}). However, starting from 
the pioneering works of Poenaru and co-workers until some recent ones, 
$E_{\nu}$ is fitted to reproduce the half-lives under consideration. 
There is no mention in these works of the fitted value being consistent 
with (\ref{vibenergy}). Such a fitting procedure performed without a 
consistency check with Eq. (\ref{vibenergy}) would be somewhat ambiguous and
outside the spirit of a proper WKB calculation. Without the condition 
(\ref{vibenergy}), $E_{\nu}$ becomes simply a parameter to compensate for 
the mismatch of the theoretical width with experiment. The interpretation 
of $E_{\nu}$ as a vibrational energy and its addition to the 
$Q$ value ($E = Q$) giving, 
$Q^{\prime} \,=\,Q\,+\,E_{\nu}$, is then not justified. The above 
point will be clarified with realistic examples of the calculation of 
half-lives of superheavy nuclei in section IV. In the next section, we 
briefly describe the potentials used for the calculations of the present work.

\section{The alpha nucleus potential}
With the objective of the present work being a critical examination of 
the various semiclassical methods used for the evaluation of alpha decay 
half-lives, we perform calculations using different available inputs in 
literature. The potential in (\ref{pot1}) is written using a double-folding 
model with realistic nucleon-nucleon interactions as given in \cite{satchler}.
The folded nuclear potential is written as,
\begin{equation}\label{potnucl}
V_n(r)\,=\,\lambda \,\int\,d{\bf r}_1\,d{\bf r}_2\,
\rho_{\alpha}({\bf r}_1) \, \rho_d({\bf r}_2)\,v({\bf r}_{12}\,=\,{\bf r}\,
+\,{\bf r}_2\,-\,{\bf r}_1,\,E)
\end{equation}
where $\rho_{\alpha}$ and $\rho_{d}$ are the densities of the alpha and the
daughter nucleus in a decay and $v({\bf r}_{12},E)$ is the nucleon-nucleon 
interaction. $|\bf{r}_{12}|$ is the distance between a nucleon in the alpha 
and a nucleon in the daughter nucleus. $v(\bf{r}_{12},E)$ is written using the 
M3Y nucleon-nucleon (NN) interaction as in \cite{satchler} as,
\begin{eqnarray}\label{nnfree}
v({\bf r}_{12},E) \,&=&\,7999\,{exp(-4\,|{\bf r}_{12}|) 
\over 4\,|{\bf r}_{12}|} 
\, -\, 2134\, {exp(-2.5\,|{\bf r}_{12}|) \over 2.5\,|{\bf r}_{12}|}\,+\,
J_{00}\,\delta({\bf r}_{12})\\ \nonumber
J_{00}\,&=&\,-276\,(1\,-\,0.005\,E_{\alpha}/A_{\alpha})\,.
\end{eqnarray}
The alpha particle 
density is given using a standard Gaussian form, namely, 
\begin{equation}\label{alphadens}
\rho_{\alpha} (r)\,=\,0.4229\,exp(-0.7024\,r^2)
\end{equation}
and the daughter nucleus density is taken to be,
\begin{equation}\label{daughter}
\rho_d(r)\,=\, {\rho_0 \over 1\,+\,exp({r - c \over a})}
\end{equation}
where $\rho_0$ is obtained by normalizing $\rho_d(r)$ to the number of 
nucleons $A_d$ and the constants are given
as $c\,=\,1.07\,A_d^{1/3}$fm and $a\,=\,0.54$fm. The equation (\ref{potnucl}) 
involves a six dimensional integral. However, the numerical evaluation becomes
simpler if one works in momentum space as shown in \cite{satchler}. 
The constant $\lambda$ is determined by imposing the Bohr-Sommerfeld 
quantization condition (\ref{bohrsomer}) using the above potential. 
The number of nodes are re-expressed as $n\, = \,(G\,-\,l)\,/2$, where 
$G$ is a global quantum number obtained from fits to data \cite{chin,buck} and
$l$ is the orbital angular momentum quantum number. We shall perform 
calculations with two possible fitted values of $G$ \cite{buck}, 
namely, $22$ and $24$.
The Coulomb potential is obtained using a similar double folding procedure
with the matter densities of the alpha and the daughter replaced by their
respective charge density distributions $\rho^c_{\alpha}$ 
and $\rho^c_{d}$. Thus, double folding the 
proton proton coulomb potential, 
\begin{equation}\label{potcol}
V_c(r)\,=\,\int\,d{\bf r}_1\,d{\bf r}_2\,
\rho^c_{\alpha}({\bf r}_1 \, \rho^c_{d}({\bf r}_2)\,{e^2 \over |{\bf r}_{12}|}
\,.
\end{equation}
The charge distributions are taken to have a similar form as the matter 
distributions, except for the fact that they are normalized to the number of
protons in the alpha and the daughter.

One could further improvise the double folding potential by taking into 
account the density dependence of the NN interaction $v({\bf r}_{12})$.
For example, in \cite{chaudhuri} a reasonably good description of elastic 
alpha-nucleus scattering data was obtained by assuming a factorized form of
the density dependence as follows:
\begin{equation}\label{nninside}
\tilde{v}({\bf r}_{12}, \rho_{\alpha}, \,\rho_d,\, E)\, =\,
C\,v({\bf r}_{12}, E)\,f(\rho_{\alpha},\, E)\,f(\rho_d, \,E)  \, ,
\end{equation}
where, $f(\rho_{X},\, E)\,=\,1\,-\,\beta\,\rho_X^{2/3}$, with $X$ being either 
the $\alpha$ or $d$. The parameters, $C$ and $\beta$ were found to be energy 
independent and $C=1.3$ and $\beta = 1.01$ fm$^2$ for the range of analysed 
data between alpha particle energies of 
100 MeV and 172 MeV. Due to the lack of much information, for the
case of super heavy nuclei, $C$ was chosen to be unity and 
$\beta = 1.6$ fm$^2$ \cite{fairy}. We note here that even if the potential 
is improvised with the density dependence of the NN interaction included, 
the Bohr-Sommerfeld condition should still be satisfied in the WKB framework.
As we shall see later, the normalization $\lambda$ in (\ref{potnucl}) is
different from unity even for the above interaction.

In order to test the applicability of the formulae (\ref{sheptop}) and 
(\ref{froformula}) at the top of the barrier, we shall examine the case of 
the $l=2$, $^8$Be resonance which decays $100\%$ into two alphas. The 
nuclear potential for the $\alpha$-$\alpha$ case is taken to be 
\cite{satchler},
\begin{equation}\label{alal}
V^{\alpha \alpha}_n(r)\,=\,-122.6225\,\,exp(-0.22\,r^2)\,{\rm MeV}\,.
\end{equation}
Since the aim of the present work is to make a comparative study and 
not fit parameters to match the theoretical widths with the experimental 
ones, the alpha particle preformation probability for all the 
calculations in this work has been taken to be unity.

\section{Results and Discussions}
The objective of the calculations performed for the superheavy nuclei is
to test the sensitivity of the results to (i) the implementation of the
Bohr-Sommerfeld quantization condition (\ref{bohrsomer}) which fixes the
strength of the potential $\lambda$ in (\ref{potnucl}), 
(ii) the `fitted' global quantum number $G$ appearing in
(\ref{bohrsomer}), (iii) the density dependence of the nucleon-nucleon 
($DD-NN$) interaction and finally 
(iv) to verify if the fitted vibrational energies 
used in literature are consistent with the theoretical Eq. (\ref{vibenergy}). 
Some recent works \cite{fairy} on superheavy elements in literature 
neglect (i), (ii) and (iv) from above. In what follows, 
we shall see that the exact values of the 
decay widths of the nuclei considered do depend strongly 
on the ingredients (i), (ii) and (iv) and hence any conclusions drawn in works
which neglect these aspects of the WKB framework would have to be treated 
with caution.
\subsubsection{Sensitivity of superheavy nuclear 
half-lives to $DD-NN$ and the Bohr-Sommerfeld
condition}
In Tables I and II, the half lives 
$t_{1/2}\,=\,{\rm ln}(2)/\Gamma(Q)$, of some currently discovered 
superheavy nuclei are shown for two different choices of the 
global quantum number $G$ of the Bohr-Sommerfeld quantization condition.  
Since all the decays discussed in Tables I and II 
take place at energies $E = Q$ which are away from the extremes of the 
barrier, we use $\Gamma(Q)\,=\,\Gamma_{TPA}(Q)$ which in turn is the 
same as evaluating $\Gamma_{DH}(Q)$ or $\Gamma_{Fr\ddot{o}man}(Q)$ as shown in 
(\ref{compare3}). Theoretically, the $Q$ value of the decay is defined 
as the difference of the masses of the parent nucleus and the sum of the
masses of the alpha and the daughter nucleus 
($Q = M_{parent}\,-\,M_{\alpha}\,-\,M_{d}$). We shall however use the 
$Q$ deduced from the measured $\alpha$-particle energies, $E_{\alpha}$, 
by applying a standard recoil correction as suggested by Perlman and 
Rasmussen \cite{perlman} and frequently used in literature. With $Z_{p}$ and 
$A_{p}$ being the charge and mass numbers respectively of the parent 
nucleus, 
\begin{equation}\label{qvalue}
Q\,=\,{A_{p} \over A_{p} \,-\,4}\,\,E_{\alpha}\,+\,(65.3\,Z_{p}^{7/5}\,-\, 
80.0\,Z_{p}^{2/5}\,)\,10^{-6}\,\,{\rm MeV}\,.
\end{equation}

The results obtained using Eq. (\ref{nnfree}) for the nucleon-nucleon (NN) 
interaction are labelled as ``free-NN" in Tables I and II. The density 
dependent NN interaction ($DD-NN$) calculations use Eq. (\ref{nninside})
instead of $v({\bf r}_{12})$ in (\ref{potnucl}) and are also shown in 
the tables.
One can see that the introduction of density dependence in the NN 
interaction reduces 
the lifetimes $t_{1/2}$ by an order of magnitude as compared to the free NN 
results.  
Neglecting the BS condition (i.e. $\lambda = 1$) however, `increases' the 
$DD-NN t_{1/2}$ by two orders of magnitude as compared to the proper $DD-NN$ 
calculation using the BS condition with $\lambda$ around 2.
Any conclusions based on calculations neglecting the BS condition 
can hence be quite misleading.
In an ideal case, when the potential $V_n(r)$ is known 
exactly for a particular system, one would expect $\lambda$ in 
(\ref{potnucl}) which gets fixed by the BS condition 
to be unity. One can however see that using the `free NN' interaction, 
the value of $\lambda$ ranges around $0.6 - 0.7$, while for the $DD-NN$ case
it is in the range of $2 - 2.3$. 
It is however interesting to note that the
value of $\lambda$ hardly depends on the mass or atomic number of the 
parent nucleus for the considered range of nuclei. 
The parameter $C$ in (\ref{nninside}) was in fact chosen to
be unity due to lack of information. One could rather choose $C=2$ 
and $C=2.3$ leading to a $\lambda$ close to 1 for the $DD-NN$ cases in 
Tables I and II respectively.

Comparing the numbers in Tables I and II, one 
observes that the half lives are reduced on increasing the value of 
the global quantum number from $G=22$ to $G=24$. 
With the primary objective of the work being a comparison of the different 
WKB approaches, the orbital quantum number $l$ was taken to be zero. Note 
however, that the introduction of the Langer modification, namely, 
$l(l+1) \rightarrow (l\,+\,1/2)^2$ for the radial one-dimensional WKB problem
introduces an additional turning point near the origin, even for the $l=0$ case
\cite{morehead}. 
This detail has also been missed out in some works \cite{fairy}.
\begin{table}
\caption{\label{tab1}Half lives in ms, for G$=22$}
\begin{ruledtabular}
\begin{tabular}{lllllll}
\hline
\textbf{Parent Nucleus}&$\lambda$\footnotemark[1] & t$_{\frac{1}{2}}$(ms)\footnotemark[1]
&$\lambda$\footnotemark[2] &t$_{\frac{1}{2}}$(ms)\footnotemark[2] &t$_{\frac{1}{2}}(ms)$\footnotemark[3] &t$_{\frac{1}{2}}(ms)$\footnotemark[4]\\
$^{271}_{106}Sg$  Q=$8.67$ MeV& $0.644$ & $17638.6$ & $2.095$ & $2794.5$ & $213197.8$& $288000$\\ \hline
$^{275}_{108}Hs$  Q=$9.44$ MeV& $0.639$ & $356.3$ & $2.080$ & $56.65$ & $4077.1$& $150$\\ \hline
$^{273}_{110}Ds$  Q=$11.368$ MeV& $0.638$ & $0.021$ & $2.072$ & $0.0034$ & $0.205$& $0.17$\\ \hline
$^{274}_{111}Rg$  Q=$11.36$ MeV& $0.639$ & $0.0439$ & $2.076$ & $0.0071$ & 
$0.46$& $6.4$\\ \hline
$^{277}112$ Q=$11.3$ MeV & $0.637$ & $0.117$ & $2.072$ & $0.019$ & $1.253$& $0.69$\\ \hline
$^{286}114$  Q=$10.35$ MeV& $0.634$ & $102.07$ & $2.067$ & $15.61$ & $1248.3$& $400$\\ \hline
$^{293}116$  Q=$10.67$ MeV& $0.627$ & $57.583$ & $2.049$ & $8.701$ & $679.6$& $53$\\ \hline
$^{294}118$  Q=$11.81$ MeV& $0.625$ & $0.416$ & $2.042$ & $0.064$ & $4.594$& $1.8$\\ \hline
\end{tabular}
\end{ruledtabular}
\footnotetext[1]{Free NN, $^b$Density dependent NN, $^c$Density 
dependent NN with $\lambda=1$, $^d$Experimental value}
\end{table}
%\vspace{10cm}
\begin{table}
\caption{\label{tab2}Half lives in ms, for G$=24$}
\begin{ruledtabular}
\begin{tabular}{lllllll}
\hline
\textbf{Parent Nucleus}&$\lambda$\footnotemark[1] & t$_{\frac{1}{2}}$(ms)\footnotemark[1]
&$\lambda$\footnotemark[2] &t$_{\frac{1}{2}}$(ms)\footnotemark[2] &t$_{\frac{1}{2}}(ms)$\footnotemark[3] &t$_{\frac{1}{2}}(ms)$\footnotemark[4]\\
$^{271}_{106}Sg$  Q=$8.67$ MeV& $0.720$ & $10511.8$ & $2.337$ & $1680.8$ & $213197.8$& $288000$\\ \hline
$^{275}_{108}Hs$  Q=$9.44$ MeV& $0.715$ & $213.1$ & $2.320$ & $34.185$ & $4077.1$& $150$\\ \hline
$^{273}_{110}Ds$  Q=$11.37$ MeV& $0.713$ & $0.012$ & $2.312$ & $0.0021$ & $0.205$& $0.17$\\ \hline
$^{274}_{111}Rg$  Q=$11.36$ MeV& $0.714$ & $0.027$ & $2.316$ & $0.0044$ & 
$0.46$& $6.4$\\ \hline
$^{277}112$ Q=$11.3$ MeV & $0.712$ & $0.0702$ & $2.311$ & $0.0114$ & $1.253$& $0.69$\\ \hline
$^{286}114$  Q=$10.35$ MeV& $0.707$ & $60.57$ & $2.302$ & $9.355$ & $1248.3$& 
$400$\\ \hline
$^{293}116$  Q=$10.67$ MeV& $0.699$ & $34.102$ & $2.280$ & $5.213$ & $679.6$& 
$53$\\ \hline
$^{294}118$  Q=$11.81$ MeV& $0.697$ & $0.248$ & $2.272$ & $0.038$ & $4.594$& $1.8$\\ \hline
\end{tabular}
\end{ruledtabular}
\footnotetext[1]{Free NN, $^b$Density dependent NN, $^c$Density 
dependent NN with $\lambda=1$, $^d$Experimental value}
\end{table}

In passing, we note that in some of the cases like the decay, 
$^{286}114 \rightarrow \,\alpha\,+\, ^{282}112$ for example, the lifetime 
of the daughter nucleus, $t_{1/2}(^{282}112) = 0.5$ ms,  
is much smaller than that of the parent, $t_{1/2}(^{286}114) = 160$ ms
(implying that the daughter in the cluster decays before the parent can decay). 
The application of quantum tunneling (which assumes a preformed cluster) 
to such a problem would be somewhat 
ambiguous with the daughter decaying faster; 
however, one could also argue that the daughter inside the 
cluster does not decay as fast as the free one and the picture is still 
valid. Indeed, in recent literature the tunneling picture is used without
considerations of the lifetimes of the parent and the daughter nuclei.

\subsubsection{Assault frequencies and vibration energies}
\begin{table}
\caption{\label{tab3}Comparison of the vibrational energies 
and corresponding 
assault frequencies obtained from fitted values, Eq. (\ref{vibtest}) and 
Eq. (\ref{vibtpa}).}
\begin{ruledtabular}
\begin{tabular}{lllllll}
\hline
\textbf{Parent}&$E_{\nu, fit}$  &$E_{\nu, test}$
&$E_{\nu, TPA}$ &$\nu_{fit} = ({2 E_{\nu, fit} \over h})$ &
$\nu_{test} = ({2 E_{\nu, test} \over h})$&
$\nu_{TPA} = ({2 E_{\nu, TPA} \over h})$\\
\textbf{Nucleus}& (MeV)  & (MeV) & (MeV) &($10^{21} s^{-1}$) 
&($10^{21} s^{-1}$)&($10^{21} s^{-1})  $
 \\
$^{271}_{106}Sg$  & 0.786 & 4.130 &6.043 &0.380  &1.998&2.923 \\ \hline
$^{275}_{108}Hs$  & 0.856 & 4.257 &5.987 &0.414  &2.059&2.896 \\ \hline
$^{273}_{110}Ds$  & 1.031 & 4.226 &5.903 &0.499  &2.044&2.856 \\ \hline
$^{274}_{111}Rg$  & 0.871 & 4.220 &5.917 &0.421  &2.041&2.863 \\ \hline
$^{277}112$  & 1.025 & 4.210 &5.913 &0.496  &2.036&2.861 \\ \hline
$^{286}114$  & 1.082 & 4.184 &5.949 &0.523  &2.023&2.878 \\ \hline
$^{293}116$  & 0.968 & 4.153 &5.910 &0.468  &2.010&2.859 \\ \hline
$^{294}118$  & 1.234 & 4.127 &5.856 &0.597  &1.996&2.833 \\ \hline
\end{tabular}
\end{ruledtabular}
\end{table}
We shall now examine the evaluation of the widths using $\Gamma_{SAFM}$ 
of Eq. (\ref{safm2}). As mentioned before, the only difference in the 
evaluation of $\Gamma_{SAFM}$ as compared to $\Gamma_{TPA}$ is in the
replacement of $Q$ by $Q\,+\,E_{\nu}$. The zero point vibration energies, 
$E_{\nu}$, are usually taken from fits \cite{poenaruzphys} and are given
for the superheavy case \cite{fairy} as, $E_{\nu}\,=\,0.1045Q$ for even-even, 
$E_{\nu}\,=\,0.0962Q$ for odd Z - even N,
$E_{\nu}\,=\,0.0907Q$ for even Z - odd N and $E_{\nu}\,=\,0.0767Q$ for 
odd Z - odd N parent nuclei. With an average $Q$ value for the superheavy 
nuclei around $Q=10$, one could say that $E_{\nu}$ would be of the order of
$1$ MeV. Such values are however not consistent with Eq. (\ref{vibenergy}). 
If for example, we provide the above values of $E_{\nu}$ 
from fits as an input for the
right hand side of Eq. (\ref{vibenergy}), the outcome (which in principle 
must be $E_{\nu}$ itself) turns out to be a much larger energy. 
For the eight nuclei considered in tables I and II, we evaluated the right 
hand side of Eq. (\ref{vibenergy}) providing $E_{\nu}$ as an input. To be 
precise, we evaluated,  
\begin{equation}\label{vibtest}
E_{\nu,test}\,=\,{\hbar^2 \,\pi \over 2\,\mu}\,\,\biggl [\,\int_{r_1}^{r_2}\, 
{dr \over \sqrt{(2\mu/\hbar^2)\,(V(r) \,-\,Q\,-\,E_{\nu, fit})}
} \,\biggr ]^{-1}\,\,.
\end{equation}
using the fitted values of $E_{\nu}$, defined as 
$E_{\nu, fit}$ in the equation above. The $E_{\nu,test}$'s 
are not the same as $E_{\nu, fit}$'s (as should have been the case due to 
Eq. (\ref{vibenergy})) and are listed in
Table III. The calculations were done with 
the density dependent NN interaction in the nuclear potential and with 
the value of $\lambda\,=\,1$ to perform a comparison with the works which
use the SAFM. The vibration energy is in fact related to the assault frequency 
at the barrier as, 
$E_{\nu} \,=\, (1/2)\,\hbar\,(2\,\pi\,\nu)$. In Table III we also list the
assault frequencies corresponding to the fitted $E_{\nu}$'s 
as used in the SAFM models and to the calculated $E_{\nu,test}$ values. 
For comparison, we present the assault frequencies appearing in the
widths, $\Gamma_{TPA}$, $\Gamma_{DH}$ and $\Gamma_{Fr\ddot{o}man}$ which are the
same in all the three cases (see Eq. (\ref{period})) 
and label them as $\nu_{TPA}$. The corresponding $E_{\nu, TPA}$ is given as,
\begin{equation}\label{vibtpa}
E_{\nu,TPA}\,=\,{\hbar^2 \,\pi \over 2\,\mu}\,\,\biggl [\,\int_{r_1}^{r_2}\, 
{dr \over \sqrt{(2\mu/\hbar^2)\,(V(r) \,-\,Q)}
} \,\biggr ]^{-1}\,\,.
\end{equation}
The above calculation of $E_{\nu,TPA}$ is performed for a density dependent 
NN interaction and including the BS condition. The assault frequencies, 
$\nu_{TPA}$ are of the order of $10^{21} s^{-1}$ 
which is more like the standard
result expected for alpha particle tunneling \cite{bookheyde}. 
\subsubsection{Improved WKB formula at the top of the barrier}
Finally, we discuss the result of the application of the improved formulae 
(\ref{sheptop}) and (\ref{froformula}) for the decay taking place with an 
energy close to the top of the barrier. Such examples are indeed difficult 
to find among the alpha decay of nuclei. A suitable one is the decay of the
$^8$Be ($2^+$) level at $3.03$ MeV above its ground state. The 
experimental width of this state is $1.513$ MeV with $100\%$ $\alpha$ decay. 
Using the analytical nuclear 
potential (\ref{alal}), the folded Coulomb potential and the 
usual formula (\ref{gurwidth}) 
for the WKB decay width $\Gamma_{TPA}$ for regions away from 
the barrier gives a theoretical width of $1.232$ MeV for this level. 
However, with the barrier height being $3.27$ MeV and the $Q$ value of the 
decay $^8$Be $\rightarrow \alpha \,+\,\alpha$ being $3.122$ MeV, the use of
the standard WKB formula (\ref{gurwidth}) is not recommendable. Instead, if 
we use Eq. (\ref{froformula}) but with a positive sign in the denominator 
as explained below (\ref{froformula}), the width 
$\Gamma^{top}_{Fr\ddot{o}man}(Q)\,=\,
1.535$ MeV and is closer to the experimental value of $1.513$ MeV. Using the
formula of \cite{shepard}, $\Gamma_{DH}^{top}(Q)\,=\,1.53$ MeV which is 
again close to the experimental number as well as 
consistent with $\Gamma^{top}_{Fr\ddot{o}man}(Q)$. If we use the expression of 
\cite{froeman} as it is in (\ref{froformula}) with a minus sign in the 
denominator, the width turns out to be $0.63$ MeV which indeed worsens the 
result of $1.232$ MeV, 
obtained with the standard WKB formula and also disagrees 
with the $\Gamma_{DH}^{top}(Q)\,=\,1.53$ MeV. Since the authors had a choice 
of the sign in \cite{froeman} 
and chose the negative sign without any particular argument, we guess that
the choice should have rather been the opposite and the expression should 
be read with a 
$T\, +\, 2\,\hbar\,{d\sigma/ dE}$ in the denominator.

\section{Conclusions}
To summarize, we first performed a survey of the available WKB 
decay width formulae in literature, which were 
obtained using different models and approaches. After having noted the 
similarities as well as differences in the various approaches, we applied them 
to the calculation of the half lives of super heavy nuclei which form a topic
of current interest. The motivation to apply for the case of superheavy nuclei 
was also to emphasize the need for performing calculations remaining 
consistently within the spirit of the WKB approximation. Following are the 
main observations of the present work:\\
(i) The decay widths of the super heavy nuclei are sensitive to the 
input of density dependence in the nucleon-nucleon interaction
of the nucleons in the $\alpha$ particle and the daughter nucleus. 
Since the half lives can reduce by an order of magnitude as compared to the
results with a free NN interaction, any conclusions drawn in such works 
regarding the angular momentum, $`l'$ values, become model dependent.\\
(ii) Conclusions obtained in some recent fission model based calculations of 
super heavy nuclei neglect the Bohr-Sommerfeld condition which 
amounts to discarding the semi classical nature inherent to 
the WKB approach. We find once again that the half lives can change by orders 
of magnitude by neglecting this condition. \\
(iii) The results as in (ii) above, often seem to be in agreement 
with data (see $t_{1/2}^c$ in Tables I and II). However, in comparing theory
with experiment one should not get tempted to choose an inconsistent 
approach for the sake of obtaining agreement as is often done in the SAFM
calculations. \\
(iv) The assault frequencies appearing in the fission model calculations 
are shown to disagree with three other approaches existing in 
literature. These frequencies and hence the vibrational energies which are 
fitted in the fission model based calculations are in principle inconsistent 
with the formulae obtained from the standard WKB method.\\
(v) Improved formulae for the decay at the top of the barrier are compared 
and applied to a realistic example. We suggest the flip of a sign in the 
denominator of the 
expression (\ref{froformula}) obtained in the work of Drukarev, Fr\"oman and 
Fr\"oman \cite{froeman}. Such a sign flip brings the results in agreement 
with experiments as well as with Eq. (\ref{sheptop}) for the width from 
another work \cite{shepard}. The sign flip is consistent with the theory in 
\cite{froeman}, since at some point in their derivation one encounters a 
choice of signs. 

To compare theoretical WKB widths with experiment, it is mandatory to 
perform a consistent calculation, taking into account carefully the details 
like the Langer modification, Bohr-Sommerfeld condition and a theoretically
derived vibrational energy. 
We conclude by mentioning that any attempt to extract physical information 
from fitted parameters (such as the alpha cluster preformation probabilities
or the unknown angular momenta of superheavy nuclei)
while calculating the half lives within the
WKB approximation should bear in mind the limitations introduced in the model 
due to the sensitivities mentioned above.

\end{document}